\documentclass[preprint2, floatfix,final]{aastex} 
\usepackage{graphicx}
\usepackage{textcomp}
\usepackage[fleqn]{amsmath}
\usepackage{mathtools}
\usepackage{natbib}

\newcommand{\V}[1]{\mathbf{#1}} 
\newcommand{\T}[1]{\texttt{#1}} 
\newcommand\Alfven{Alfv\'en }
\newcommand\Alfvenic{Alfv\'enic }
\newcommand{\figref}[1]{Figure~\ref{#1}}

\newcommand{\xhat}{\mbox{$\hat{\mathbf{x}}$}} 
\newcommand{\yhat}{\mbox{$\hat{\mathbf{y}}$}} 
\newcommand{\zhat}{\mbox{$\hat{\mathbf{z}}$}}

\usepackage{color}
\usepackage{xcolor}
\begin{document}


\title{Energy Dissipation and Landau Damping \\in Two- and Three-Dimensional Plasma Turbulence}

\author{Tak~Chu~Li\altaffilmark{1}, Gregory~G. Howes\altaffilmark{1}, Kristopher~G.~Klein\altaffilmark{2} and Jason~M.~TenBarge\altaffilmark{3}}
\affil{\altaffilmark{1}Department of Physics and Astronomy, University of Iowa, Iowa City, Iowa 52242, USA}
\affil{\altaffilmark{2}Space Science Center, University of New Hampshire, Durham, New Hampshire 03824, USA}
\affil{\altaffilmark{3}IREAP, University of Maryland, College Park, Maryland 20742, USA}


\begin{abstract}
Plasma turbulence is ubiquitous in space and astrophysical plasmas,
playing an important role in plasma energization, but the physical
mechanisms leading to dissipation of the turbulent energy remain to
be definitively identified. Kinetic simulations in two dimensions (2D) have been extensively used to study the dissipation process. How the limitation to 2D affects energy dissipation remains unclear.
This work provides a model of comparison between two- and
three-dimensional (3D) plasma turbulence using gyrokinetic simulations; it also explores the dynamics of distribution functions during the dissipation process. It is found that both 2D and 3D nonlinear gyrokinetic simulations of a low-beta plasma generate electron velocity-space structures with the same characteristics as that of linear Landau damping
of \Alfven waves in a 3D linear simulation. The continual occurrence of the velocity-space structures throughout the turbulence simulations suggests that the action of Landau damping may be responsible for the turbulent energy transfer to electrons in both 2D and 3D, and makes
possible the subsequent irreversible heating of the
  plasma through collisional smoothing of the velocity-space
  fluctuations. Although, in the 2D case where variation along the
equilibrium magnetic field is absent, it may be expected that Landau damping is not possible, a common trigonometric factor appears in the 2D resonant denominator, leaving the resonance condition unchanged from the 3D case. The evolution of the 2D and 3D
cases is qualitatively similar. However, quantitatively the nonlinear energy cascade and subsequent dissipation is significantly slower in the 2D case.
\end{abstract}



\section{Introduction}

In a wide variety of space and astrophysical systems such as the solar corona, solar wind, the interstellar medium and galaxy clusters, plasma turbulence plays a governing role in the transfer of energy from large-scale motions down to small scales, where that
energy is ultimately converted to plasma heat. A fundamental question
at the frontier of astrophysics is what physical mechanisms determine
the dissipation of the turbulence, specifically how the energy of the
turbulent fluctuations in a weakly collisional plasma is ultimately
converted to heat or other forms of particle energization.

To investigate numerically the dynamics of the collisionless
wave-particle interactions that lead to damping of the turbulent
electromagnetic fluctuations requires kinetic simulations, but their
high-dimensionality requires significant computational power that
often limits numerical studies to only two spatial dimensions \citep{Gary:2008,Parashar:2009,Servidio:2012,Verscharen:2012a,Wan:2012,Markovskii:2013,Perrone:2013,Wu:2013a,Wu:2013b,Che:2014,Haynes:2014,Narita:2014,Parashar:2014b,Franci:2015}. How
this limitation to 2D constrains the available energy dissipation
pathways remains an open question
\citep{Howes:2015a,Wan:2015,Servidio:2015}.

This work aims at providing a model of comparison between 2D and 3D turbulence systems in a low-$\beta$ plasma, focusing on energy dissipation, and exploring the dynamics of distribution functions as a novel means to characterize the energy transfer mechanism in velocity space. 

In the following, we present results from gyrokinetic simulations showing quantitative difference in the energy dissipation of 2D and 3D nonlinear turbulence simulations and the development of velocity-space structures that share the same characteristics as that of linear Landau damping of \Alfven waves in a 3D linear simulation. The velocity-space structures are observed throughout the entire evolution of both 2D and 3D turbulence simulations, implying the continual occurrence of Landau damping during the dissipation of turbulence in both systems. Although, in the 2D case with no
variation along the equilibrium magnetic field, it may be expected
that Landau damping is not possible, we explain why the limitation to
2D does not prohibit Landau damping.  This 2D limitation does,
however, alter the quantitative evolution of the energy, likely due to
a less rapid nonlinear cascade of energy to small scales compared to
the 3D case.

\section{Simulation Code}
The Astrophysical Gyrokinetics Code, or \T{AstroGK},
described in detail in Ref.~\citep{Numata:2010}, evolves the perturbed
gyroaveraged distribution function $h_s(x,y,z,\lambda,\varepsilon)$
for each species $s$, the scalar potential $\varphi$, parallel vector
potential $A_\parallel$, and the parallel magnetic field perturbation
$\delta B_\parallel$ according to the gyrokinetic equation and the
gyroaveraged Maxwell's equations \citep{Frieman:1982,Howes:2006}, where
$\parallel$ is along the total local magnetic field $\V{B}=B_0
\zhat+ \delta \V{B}$. The velocity-space coordinates are
$\lambda=v_\perp^2/v^2$ and $\varepsilon=v^2/2$. The domain is a
periodic box of size $L_{\perp }^2 \times L_z$, elongated along the
equilibrium magnetic field, $\V{B}_0=B_0 \zhat$. Note that, in the
gyrokinetic formalism, all quantities may be rescaled to any parallel
dimension satisfying $\epsilon \equiv L_\perp /L_z \ll 1$. Uniform
Maxwellian equilibria for ions (protons) and electrons are
used. Spatial dimensions $(x,y)$ perpendicular to the equilibrium
field are treated pseudospectrally; an upwind finite-difference scheme
is used in the parallel direction. Collisions are incorporated using a
fully conservative, linearized Landau collision
operator that includes energy diffusion and pitch-angle scattering
due to electron-electron, ion-ion, and electron-ion
  collisions \citep{Abel:2008,Barnes:2009}, yielding an isotropic
  Maxwellian stationary solution.
 
\section{Diagnostics}
Two diagnostics are used to explore the
energetics and velocity-space structure in the 2D and 3D turbulence
simulations. The \emph{energy diagnostic} examines the partitioning of
energy among the terms of the fluctuating energy in the
simulations. Consider a kinetic plasma with each species distribution
function separated into an equilibrium and a fluctuating part,
$f_s=F_{0s} + \delta f_s$.  \T{AstroGK} evolves the perturbed
distribution function, $\delta f_s$, making it possible to follow the
exchange of energy within the \emph{total fluctuating energy} \citep{Howes:2006,Brizard:2007,Schekochihin:2009} given by
\begin{eqnarray}
\lefteqn{\delta W =} \\&& 
\int\!d^3\V{r} \left[\frac{|\delta \V{B}|^2+|\delta \V{E}|^2}{8\pi}+\sum\limits_s \left( \frac{1}{2}n_{0s}m_s|\delta
  \V{u_s}|^2 + \frac{3}{2} \delta P_s  \right) \right], \nonumber 
\label{eq:general_W}
\end{eqnarray}
where $s$ is the species index representing ions or electrons in each variable, $n_{0s}$ the equilibrium density, $m_s$ mass and $\delta\V{u_s}$ the fluctuating bulk velocity; the
\emph{non-thermal energy} in the distribution function (minus
bulk kinetic energy) is defined by $E^{(nt)}_s\equiv \int d^3 \V{r}\frac{3}{2} \delta P_s \equiv
\int d^3 \V{r} (\int d^3 \V{v}\ T_{0s}\delta f_s^2/2F_{0s} -
\frac{1}{2}n_{0s}m_s|\delta \V{u_s}|^2$) \citep{TenBarge:2014b}, where $T_{0s}$ is the equilibrium temperature. The \emph{turbulent energy} is defined as the sum of the
electromagnetic field and the bulk kinetic energies
\citep{Howes:2015b}, $E^{(turb)}\equiv \int d^3 \V{r} [ (|\delta
\V{B}|^2+|\delta \V{E}|^2)/8\pi + \sum_s \frac{1}{2}n_{0s}m_s|\delta
\V{u_s}|^2 ]$. This sum comprises the turbulent
energy because the linear terms in the evolution equations
(primarily the effect of magnetic tension) lead to an oscillatory
sloshing of energy between magnetic and bulk kinetic energies. Note that $\delta W$ includes neither the equilibrium
thermal energy, $\frac{3}{2} n_{0s} T_{0s} = \int d^3 \V{r} \int d^3 \V{v} \frac{1}{2} m_sv^2 F_{0s}$, nor the equilibrium magnetic field
energy, $\int d^3 \V{r} \ B_0^2/8\pi$. Thus, the terms of $\delta W$
represent the perturbed electromagnetic field energies, and the species
bulk kinetic and non-thermal energies.

The flow of energy in our kinetic turbulence simulations follows a
simple path.  As the system evolves, first, nonlinear interactions
transfer the turbulent energy from large to small spatial
scales. When the energy has reached sufficiently small spatial
scales, collisionless wave-particle interactions transfer energy from
the electromagnetic fields to non-thermal energy in the particle
distributions. This non-thermal energy in $\delta f_s$ manifests as
small-scale structures in the velocity space, which are ultimately
smoothed out by collisions, irreversibly converting the non-thermal
energy into thermal energy, and thereby increasing the entropy of the
plasma \citep{Howes:2006}. In \T{AstroGK}, the effect of collisions is
to remove energy from $\delta W$.  The collisional energy lost from
$\delta W$, denoted $E^{(coll)}_s$, is tracked by the energy
diagnostic, representing thermal heating of the species, but this
energy is not fed back into the equilibrium thermal temperature,
$T_{0s}$ \citep{Howes:2006,Numata:2010}.

The \emph{velocity-space diagnostic} probes structures in velocity
space that arise from the collisionless damping of the turbulent
fluctuations.  To first-order in gyrokinetic theory \citep{Howes:2006},
the distribution function is given by
\begin{equation}
f_s(v_\parallel,v_\perp) =
\left(1-\frac{q_s\varphi}{T_{0s}}\right)F_{0s}(v)+
h_s(v_\parallel,v_\perp),
\label{eq:f_e}
\end{equation} 
where $F_{0s} = (n_{0s}/\pi^{3/2}v_{ts}^3)\exp(-v^2/v_{ts}^2)$ is the
equilibrium Maxwellian distribution, $q_s\varphi/T_{0s}$ is the
Boltzmann term ($q_s$ the species charge and $\varphi$ the electric potential), $h_s$ is the first-order gyroaveraged part of the
perturbed distribution.
The complementary distribution function \citep{Schekochihin:2009},
\begin{equation}
 g_s(v_\parallel,v_\perp) = h_s(v_\parallel,v_\perp) - \frac{q_s F_{0s}}{T_{0s}}\left\langle \varphi - \frac{\V{v}_\perp\cdot\V{A}_\perp}{c} \right\rangle_{\V{R}_s},
\label{eq:g_s}
\end{equation} 
where $\langle ... \rangle$ represents gyroaveraging at constant guiding center $\V{R}_s$, removes the effect of the perpendicular bulk flow of MHD \Alfven waves, allowing kinetic dynamics of collisionless damping to be more clearly seen.

\section{Simulations}
The 2D Orszag-Tang Vortex (OTV) problem
\citep{Orszag:1979}, and various 3D generalizations, have been widely
used to study plasma turbulence
\citep{Politano:1989,Dahlburg:1989,Picone:1991,Politano:1995b,Grauer:2000,Mininni:2006,Parashar:2009,Parashar:2014b}. We
specify here a particular 3D formulation of the initial conditions
(denoted \emph{OTV3D}), given in Els\"asser variables,
$\mathbf{z}^\pm\equiv\delta\V{u} \pm \delta\V{B}/\sqrt{4\pi\rho_0}$,
by
\begin{equation}
\begin{aligned}
    \frac{\mathbf{z}^+_1}{v_A} &= -\frac{2\mbox{z}_0}{v_A}\, \hat{\mathbf{x}}\,\sin{(k_\perp y - k_\parallel z)},\,\,\,\frac{\mathbf{z}^-_1}{v_A} =  0\\
 \frac{\mathbf{z}^\pm_2}{v_A} &= \, \frac{\mbox{z}_0}{v_A}\, \hat{\mathbf{y}}\,\sin{(k_\perp x \mp k_\parallel z)}\\
 \frac{\mathbf{z}^\pm_3}{v_A} &=  \pm\frac{\mbox{z}_0}{v_A}\, \hat{\mathbf{y}}\,\sin{(2k_\perp x \mp k_\parallel z)}\\
\end{aligned}
\label{eq:elsass_3D}
\end{equation}
where $v_A = B_0/\sqrt{4\pi\rho_0}$ is the \Alfven speed,
$\rho_0=m_in_0$ is the ion mass density, $\delta \V{u}$ and $\delta \V{B}$
are perturbations in the ion bulk velocity and the magnetic field,
and $k_\perp=2\pi/L_\perp$ and $k_\parallel=2\pi/L_z$ are positive
constants.  This 3D generalization consists of counterpropagating
\Alfven waves along $B_0\hat{\mathbf{z}}$. On the mid-plane ($z=0$),
the 3D formulation reduces to the familiar 2D OTV setup
(\emph{OTV2D}), given by
\begin{equation}
\begin{aligned}
  \delta\V{u} &= \delta u [ - \sin(k_\perp y) \xhat + \sin(k_\perp x)\yhat]\\
  \delta\V{B} &= \delta B [ - \sin(k_\perp y)\xhat +  \sin(2k_\perp x) \yhat ],\\
\end{aligned}
\label{eq:vB2D}
\end{equation}
where $\mbox{z}_0= \delta u =\delta B/\sqrt{4 \pi \rho_0}$.

To resolve the kinetic mechanisms mediating the transfer of
turbulent energy to plasma energy, it is necessary to follow the
turbulent cascade from the inertial range ($k_\perp\rho_i \ll 1$) to below the electron scales ($k_\perp\rho_e >1$)
\citep{TenBarge:2013a,TenBarge:2013b,TenBarge:2014b}. Therefore, we
specify a reduced mass ratio, $m_i/m_e=25$, which, in a simulation domain of $L_{\perp}=8\pi \rho_i$ and dimensions $(n_x,n_y,n_z,n_\lambda,n_\varepsilon,n_s)=(128,128,32,64,32,2)$ in 3D (with $n_z=2$ in 2D), enables us to resolve a dynamic
range of $0.25 \le k_\perp\rho_i \le 10.5$, or $0.05 \le
k_\perp\rho_e \le 2.1$.

Plasma parameters are ion plasma beta $\beta_i = 8 \pi n_i T_{0i}/B_0^2
=0.01$ and $T_{0i}/T_{0e}=1$.  Under $\beta_i \ll 1$ conditions, the ion
dynamics are expected to contribute negligibly to the collisionless
damping via the Landau resonance. Collision frequencies of $\nu_i$ =
10$^{-5} \omega_{A0}$ and $\nu_e$ = 0.05 $\omega_{A0}$ (where
$\omega_{A0} \equiv k_{\parallel}v_A$ is a characteristic \Alfven wave
frequency in 3D) are sufficient to keep velocity space well resolved
\citep{Howes:2008a,Howes:2011a} and enable irreversible heating of the
plasma species \citep{Howes:2006}. We choose an initial amplitude that
yields a nonlinearity parameter $\chi = k_\perp
\mbox{z}_0/(k_\parallel v_A) = 1$, corresponding to \emph{critical
  balance} \citep{Goldreich:1995}, or a state of strong turbulence; the
same amplitude $\mbox{z}_0$ is used in the 2D run. Time is normalized
by the domain turnaround time $\tau_0=L_{\perp}/\mbox{z}_0$ and
velocity by the species thermal speed $v_{ts}=\sqrt{2T_{0s}/m_s}$.

Note that, for the reduced mass ratio, the low beta conditions satisfy
$\beta_i<m_e/m_i$, so the \Alfven wave transitions to an inertial
\Alfven wave \citep{Thompson:1996} at small scales. In 3D, the initial counterpropagating
\Alfven waves at the perpendicular domain scale $L_\perp$ have a
frequency of $\omega$ = 0.93$\omega_{A0}$, and the frequency drops
slowly as the perpendicular scale decreases. Simulations with
$\beta_i$=0.1, in which the \Alfven wave transitions instead to a
kinetic \Alfven wave at small scales, show results similar to those
presented here, so we believe the physical mechanism reported here is
robust for low-$\beta$ plasmas.

\begin{figure*}[t]
\centering
\hbox{ \hfill \resizebox{2.2in}{!}{\includegraphics{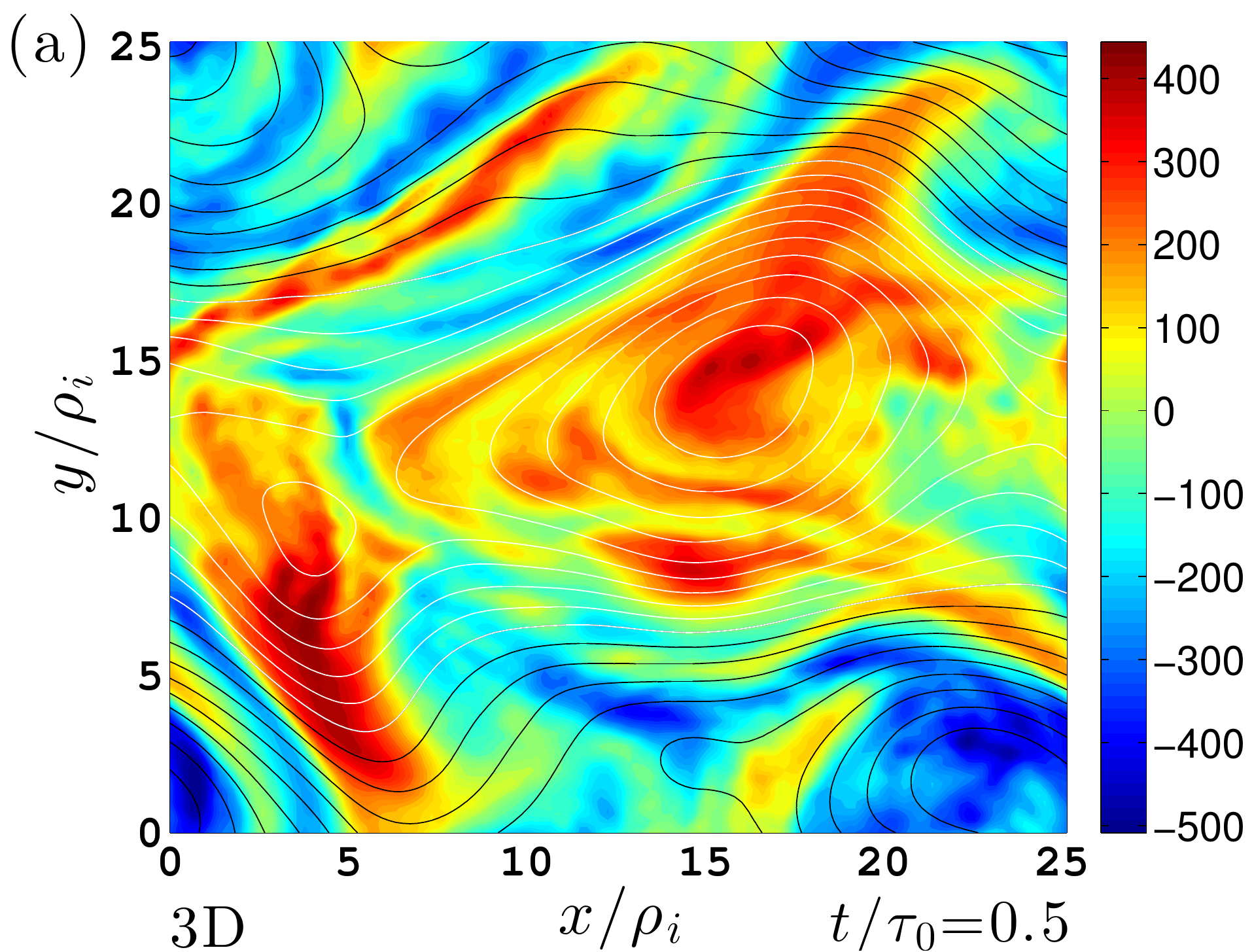}} \hfill 
\resizebox{2.2in}{!}{\includegraphics[scale=0.25]{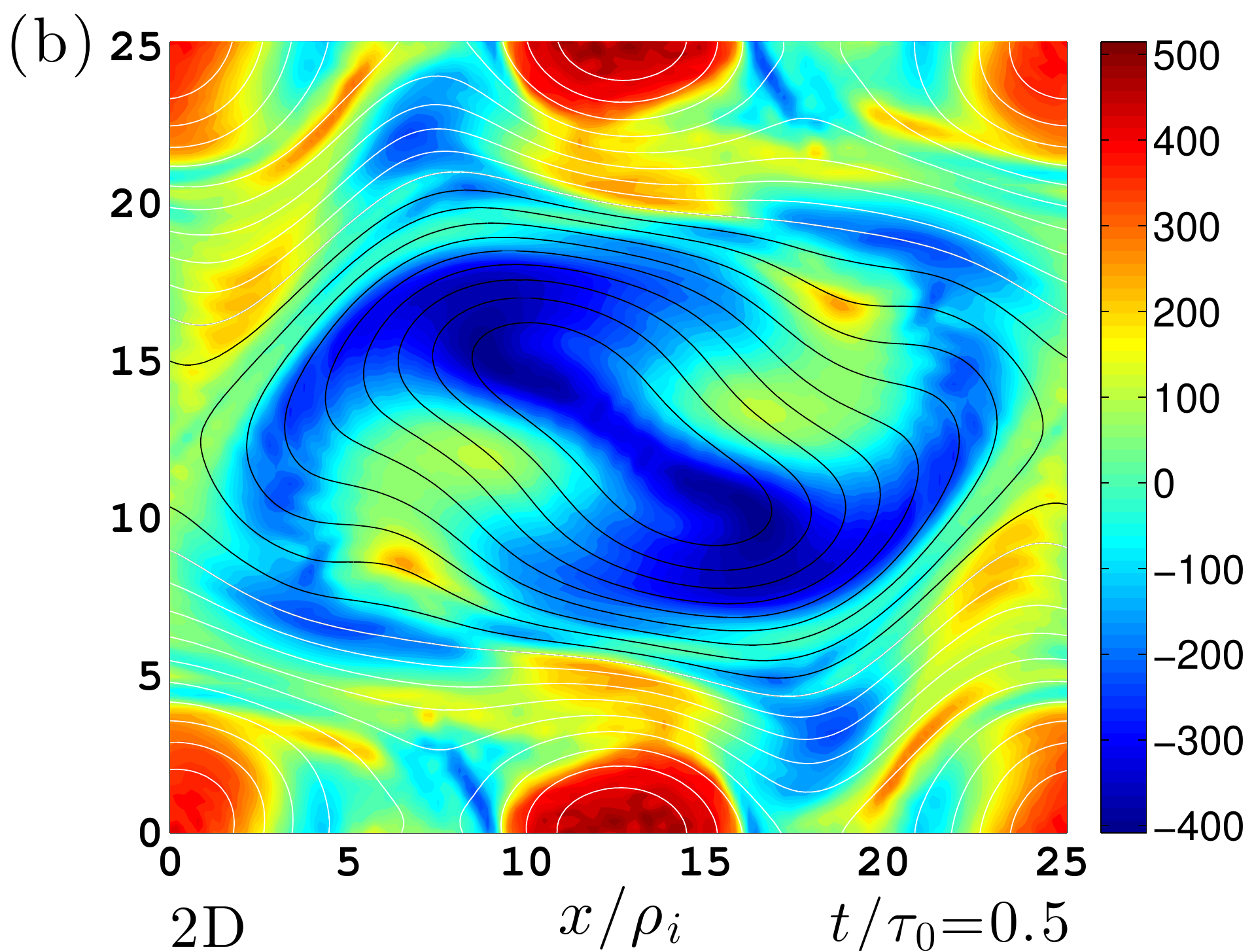}} \hfill }
\hbox{ \resizebox{2.2in}{!}{\includegraphics[scale=0.25]{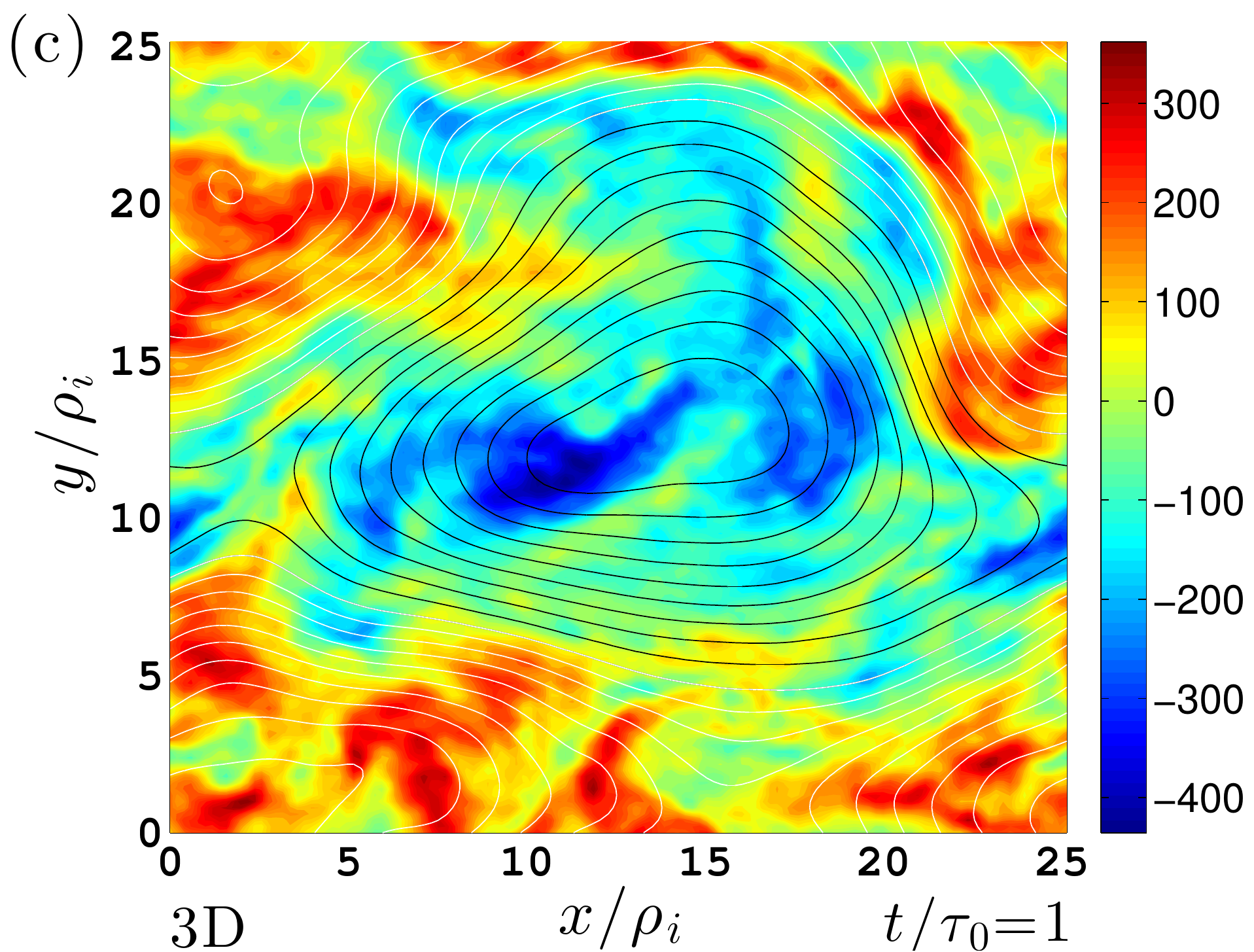}} \hfill
\resizebox{2.2in}{!}{\includegraphics[scale=0.25]{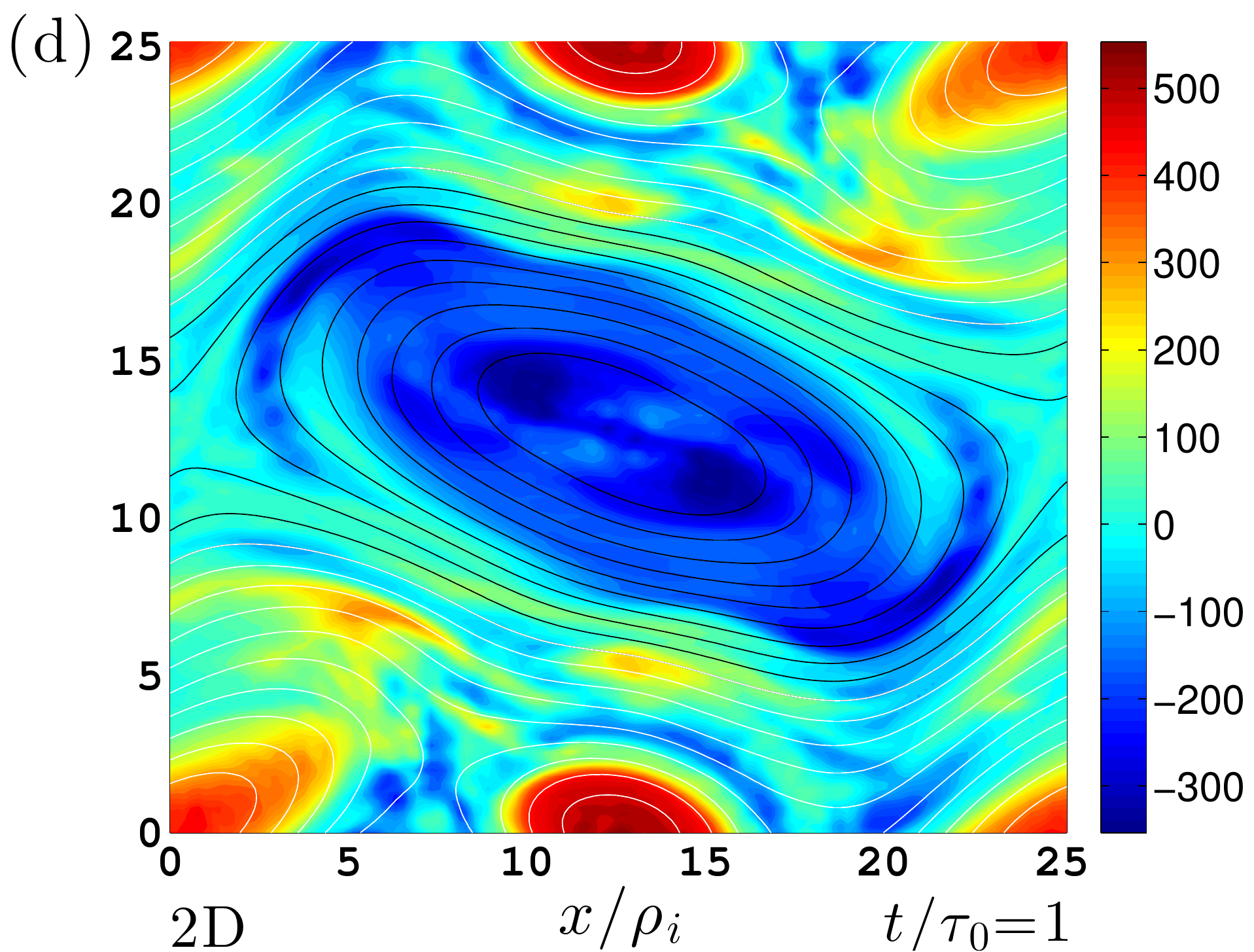}} \hfill }
\caption{ \label{fig:os23otv} Spatial profile of $J_\parallel$ (color) and
  $A_\parallel$ (contours) on the $z=0$ plane of the OTV3D (left) and
  OTV2D (right) simulations at $t/\tau_0=0.5$ (top) and $t/\tau_0=1$
  (bottom). Contours represent positive (white) and negative (black)
  values of $A_\parallel$. }
\end{figure*}

\section{Results}
\subsection{Evolution of the OTV}The OTV setup consists of an initial
flow and current pattern, which breaks up into turbulence before
$t=\tau_0$. We plot in \figref{fig:os23otv} the spatial profile of
$J_\parallel$ (color) and $A_\parallel$ (contours) from the OTV2D and the
$z=0$ plane of the OTV3D simulations. At $t=0.5\tau_0$, the initial
double vortices near the center have merged completely in both cases,
and thin current sheets have developed.  Note that $t=0.5\tau_0$ is
roughly half of the \Alfven wave period, so the $J_\parallel$ pattern at $z=0$
in 3D appears $\pi$ \textit{out of phase} (opposite colors) relative
to the 2D case.  At $t=\tau_0$, $J_\parallel$ in 3D returns to a configuration
similar to the 2D case: negative (blue) current in the center
surrounded by positive (red) currents.  The nonlinear cascade of
energy to small scales is more rapid in the 3D case, showing more
energy in small-scale magnetic structures in \figref{fig:os23otv};
this is supported by comparing the magnetic energy spectra (not
shown).


\begin{figure}[t]
\centering
\hspace{2.em}%
\hbox{ \hfill \resizebox{3.in}{!}{\includegraphics{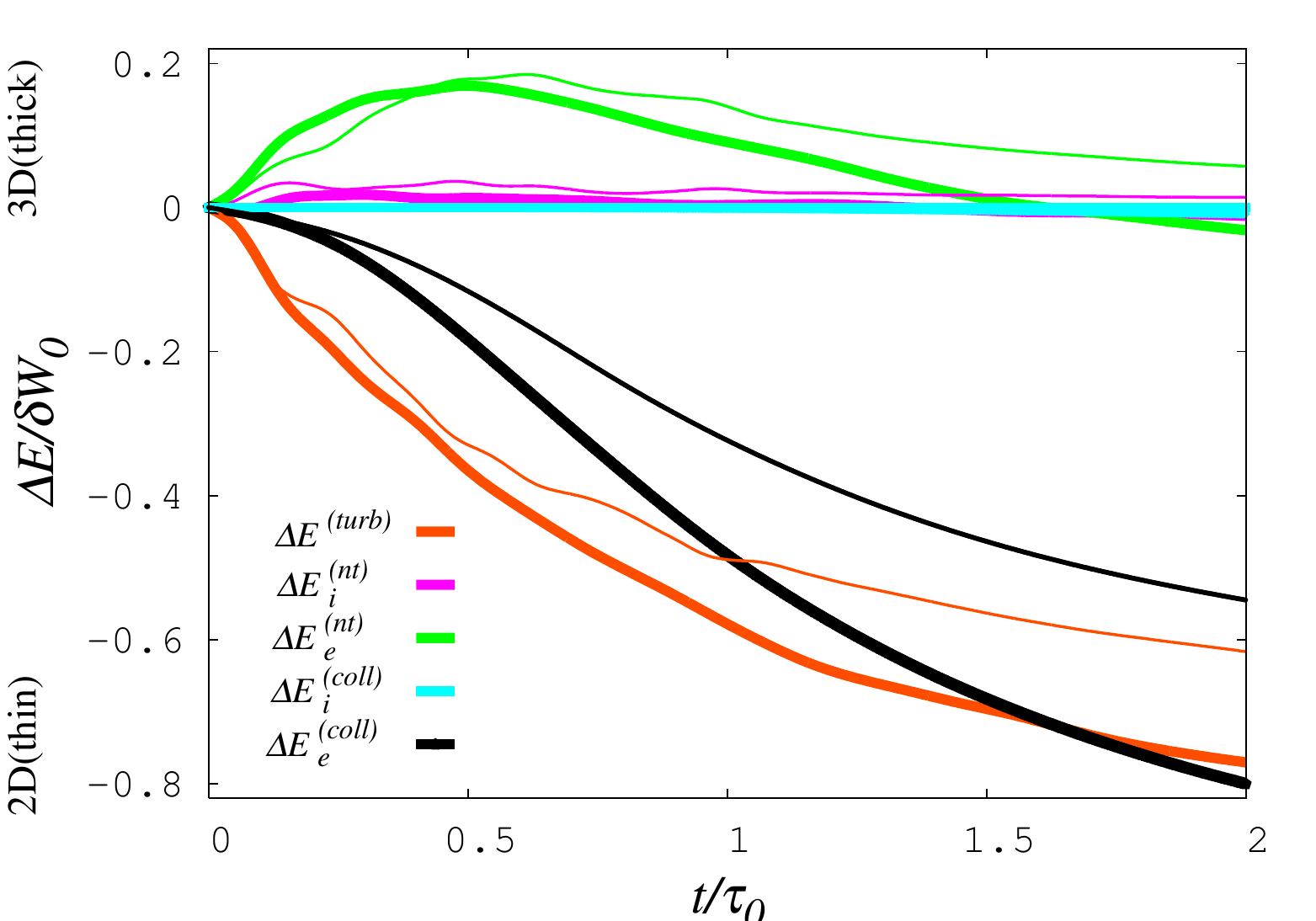}} \hfill }

\caption{Change of energy over total initial fluctuating energy,
  $\Delta E/\delta W_0$, for the turbulent energy $E^{(turb)}$ (orange),
  the non-thermal energy $E^{(nt)}_s$ of ions (magenta) and electrons
  (green), the collisionally dissipated energy $E^{(coll)}_s$ for ions
  (cyan) and electrons (black). Line thickness indicates OTV3D (thick)
  or OTV2D (thin) simulations.  \label{fig:sdk2_001turb}}
\end{figure}

\subsection{Evolution of Energy}
Next we consider the flow of energy from
turbulent fluctuations to plasma heat.  It is worthwhile mentioning
the dominant components of the turbulent energy in our
simulations. For low-frequency electromagnetic waves, $|\delta
\V{E}|^2/|\delta \V{B}|^2 \sim \mathcal{O}(v_A^2/c^2)$, so the
electric field energy is negligible in the non-relativistic limit
\citep{Howes:2006,Howes:2014a}.  Throughout the evolution, the
turbulent energy $E^{(turb)}$ is dominated by the perpendicular
magnetic and perpendicular ion bulk kinetic energy (more than
75\%). The parallel electron bulk kinetic energy, supporting the
currents in the simulation, contributes less than 20\% of the energy,
but this is artificially large due to the small mass ratio chosen,
$m_i/m_e=25$; for a realistic mass ratio of $1836$, it is not expected
to contribute significantly to the energy budget.  Finally, the
parallel magnetic, parallel ion bulk kinetic, and perpendicular
electron bulk kinetic energies contribute less than 7\% of the
turbulent energy.

\figref{fig:sdk2_001turb} shows the change in energy $\Delta E$,
relative to the initial total fluctuating energy $\delta W_0$, for the
turbulent $E^{(turb)}$ (orange), non-thermal $E^{(nt)}_s$ (magenta,
green), and collisionally dissipated $E^{(coll)}_s$ (cyan, black)
energies, where line thickness indicates OTV3D (thick) or OTV2D (thin)
simulations. Initially, turbulent energy (orange) is dominantly
transferred into electron non-thermal energy (green).  Physically,
this occurs through a two-step process: (1) the turbulent energy is
transferred nonlinearly to small spatial scales; and (2) at these
small scales, collisionless wave-particle interactions transfer energy
from the electromagnetic fields to the electrons as non-thermal energy
in the velocity distribution, $E^{(nt)}_e$.  At $t>0.25 \tau_0$, the
electron non-thermal energy begins to be significantly collisionally
dissipated (black), indicating the ultimate thermalization of electron
energy (which \T{AstroGK} removes from $\delta W$). As expected, ions
play a negligible role in the dissipation of the turbulent energy
at $\beta_i \ll 1$.

The qualitative evolution of the 2D case is remarkably similar to the
3D case, suggesting that the same kinetic physical mechanisms mediate
the dissipation in both cases. Despite this qualitative similarity,
quantitatively, the dissipation rate is significantly faster in 3D. By
$t=2\tau_0$, the 3D simulation has dissipated 80\% of $\delta W_0$,
indicating a strong turbulent cascade, but the 2D simulation has
dissipated only 55\%.

\begin{figure*}[ht] \centering
\hbox{ \hfill \resizebox{1.8in}{!}{\includegraphics[scale=0.29]{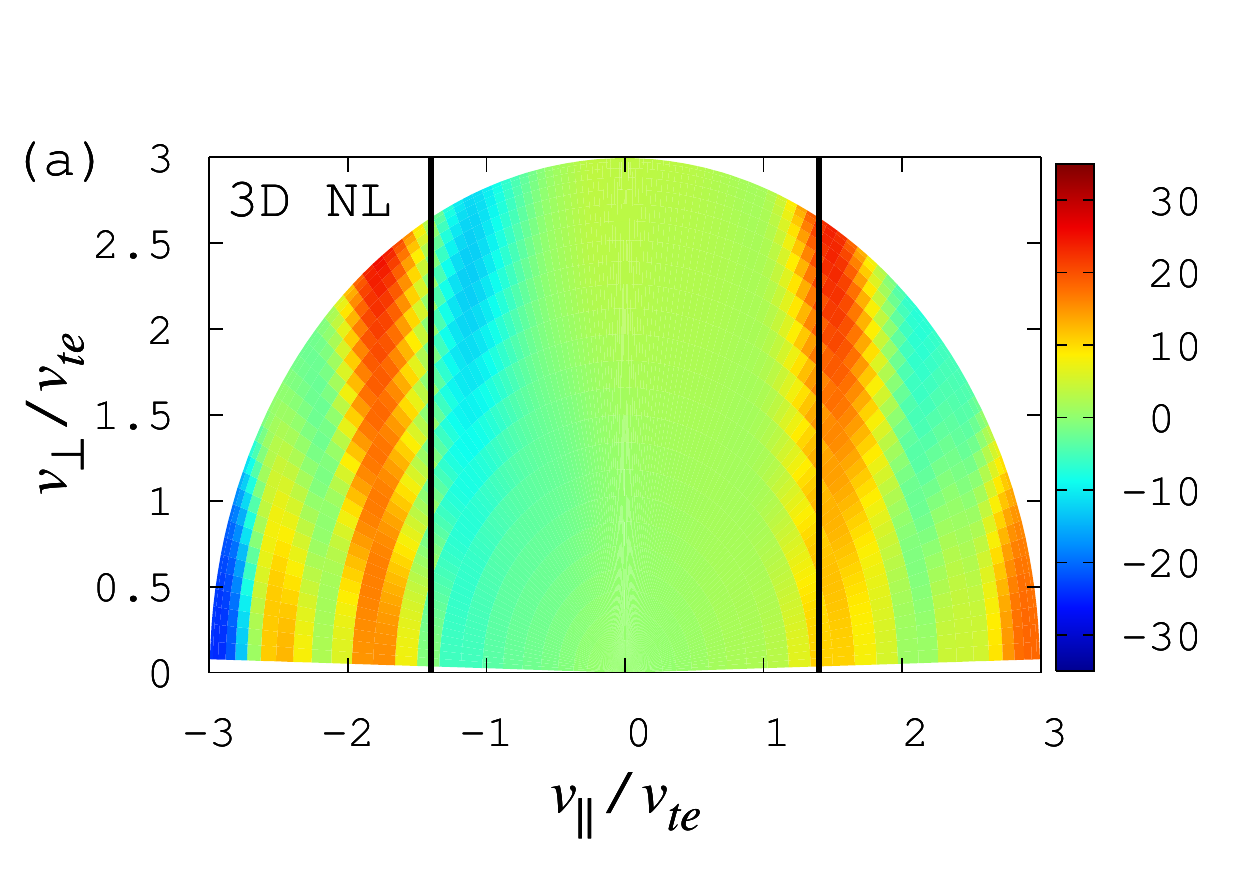}} \hfill 
  \resizebox{1.8in}{!}{\includegraphics[scale=0.29]{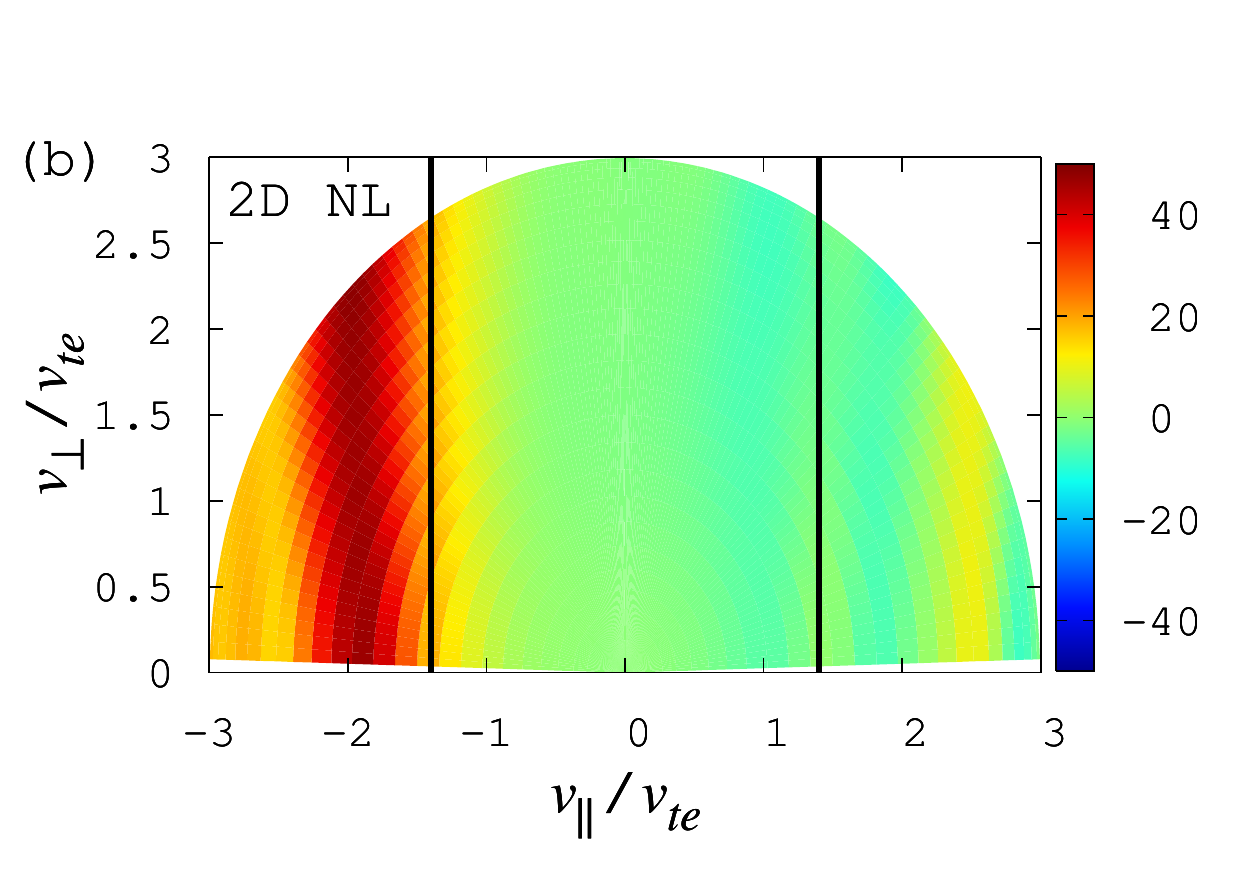}} \hfill
  \resizebox{1.8in}{!}{\includegraphics[scale=0.29]{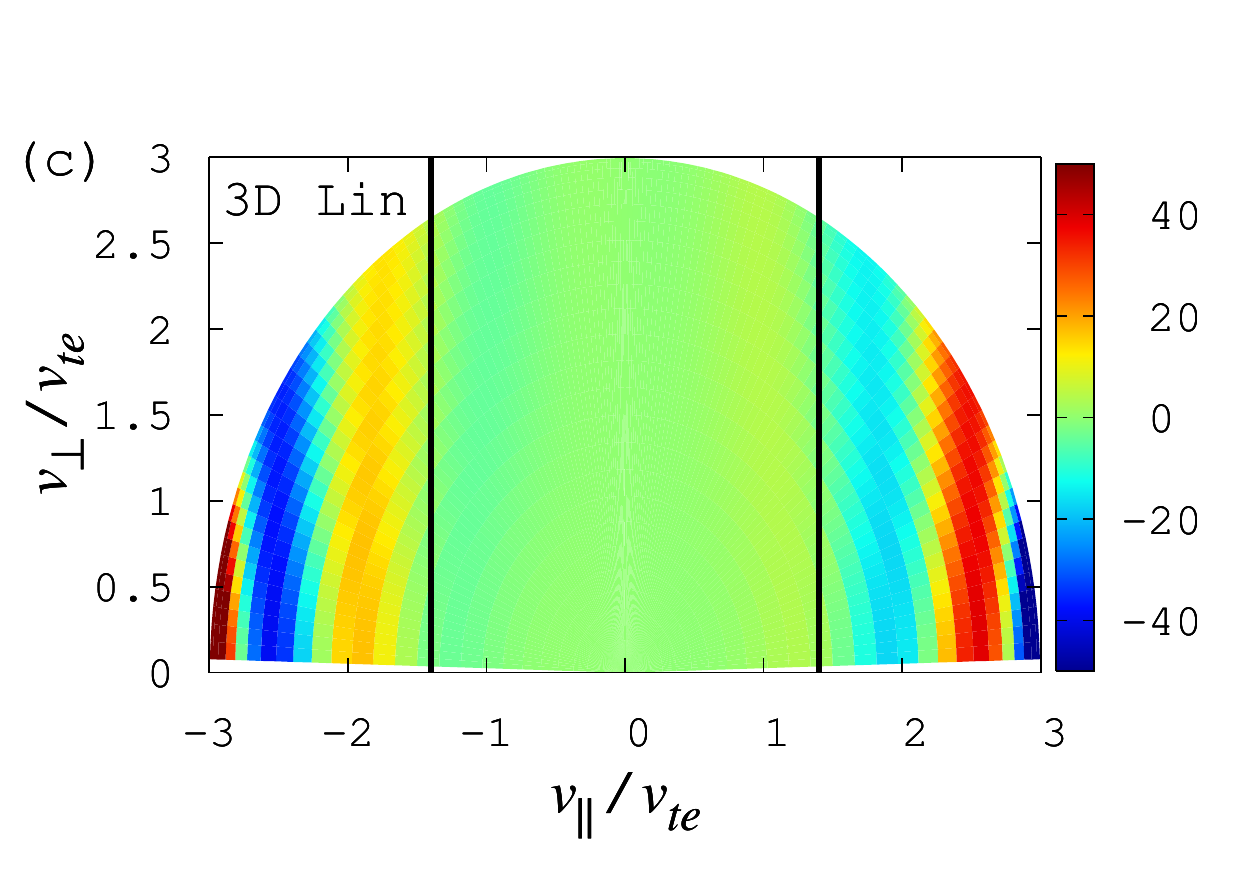}} \hfill  
  \resizebox{1.8in}{!}{\includegraphics[scale=0.29]{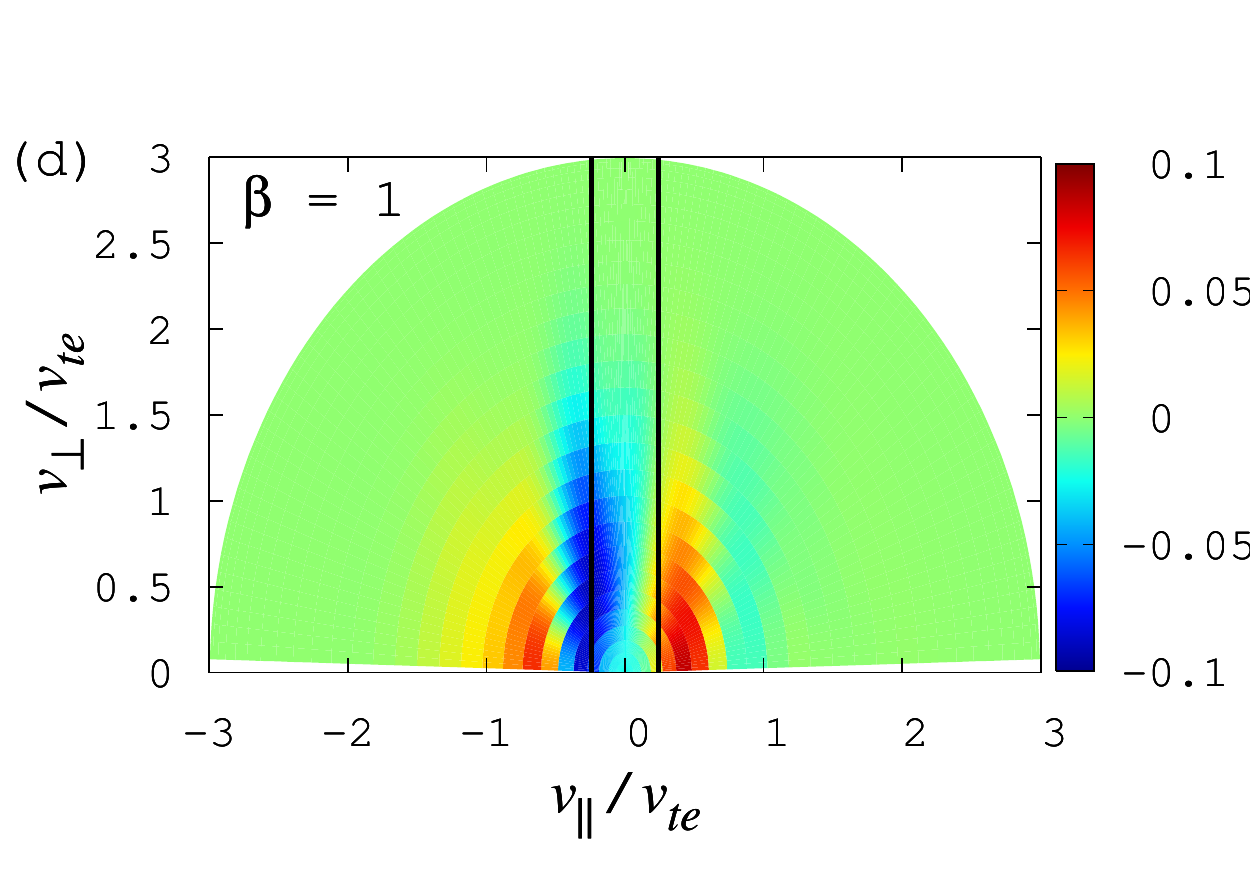}} \hfill }
\hbox{ \hfill    \resizebox{1.8in}{!}{\includegraphics[scale=0.29]{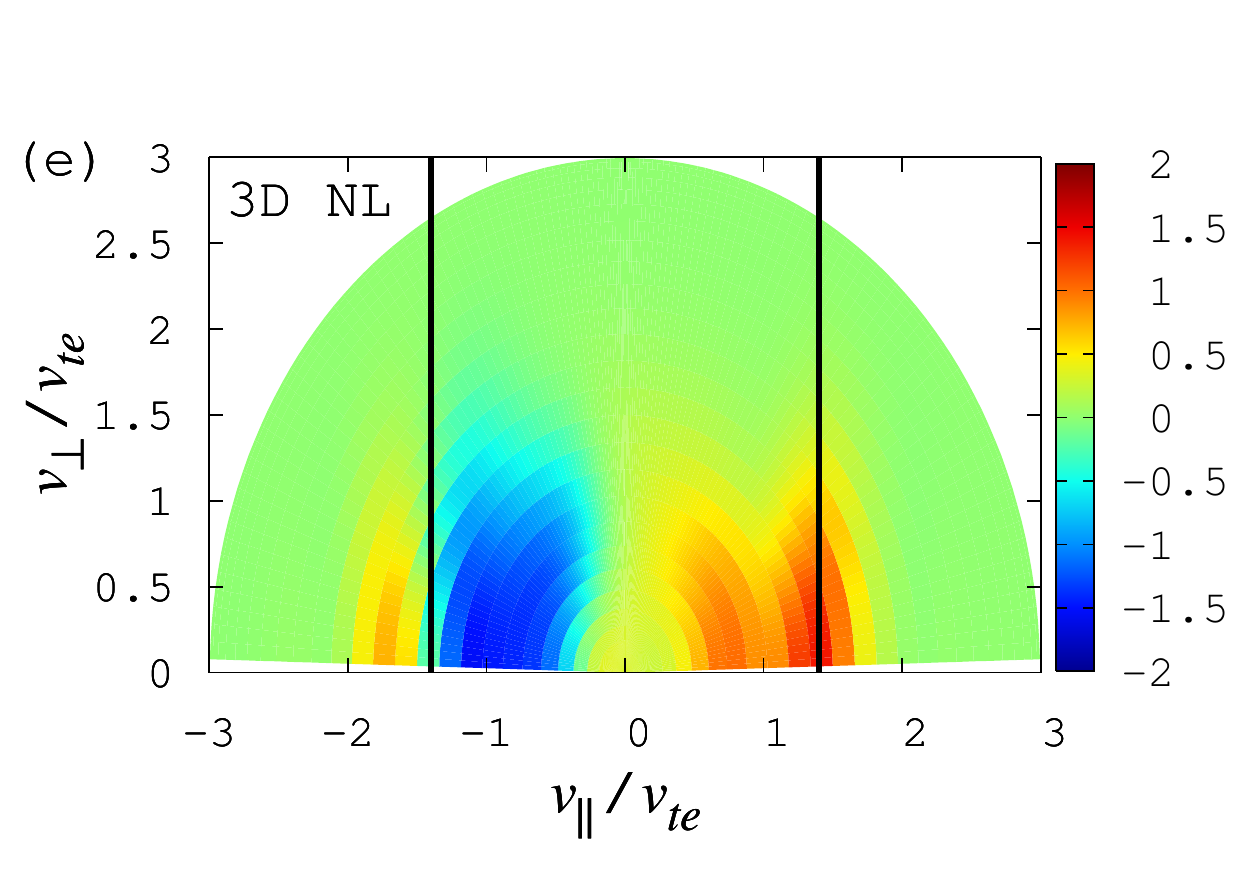}} \hfill
  \resizebox{1.8in}{!}{\includegraphics[scale=0.29]{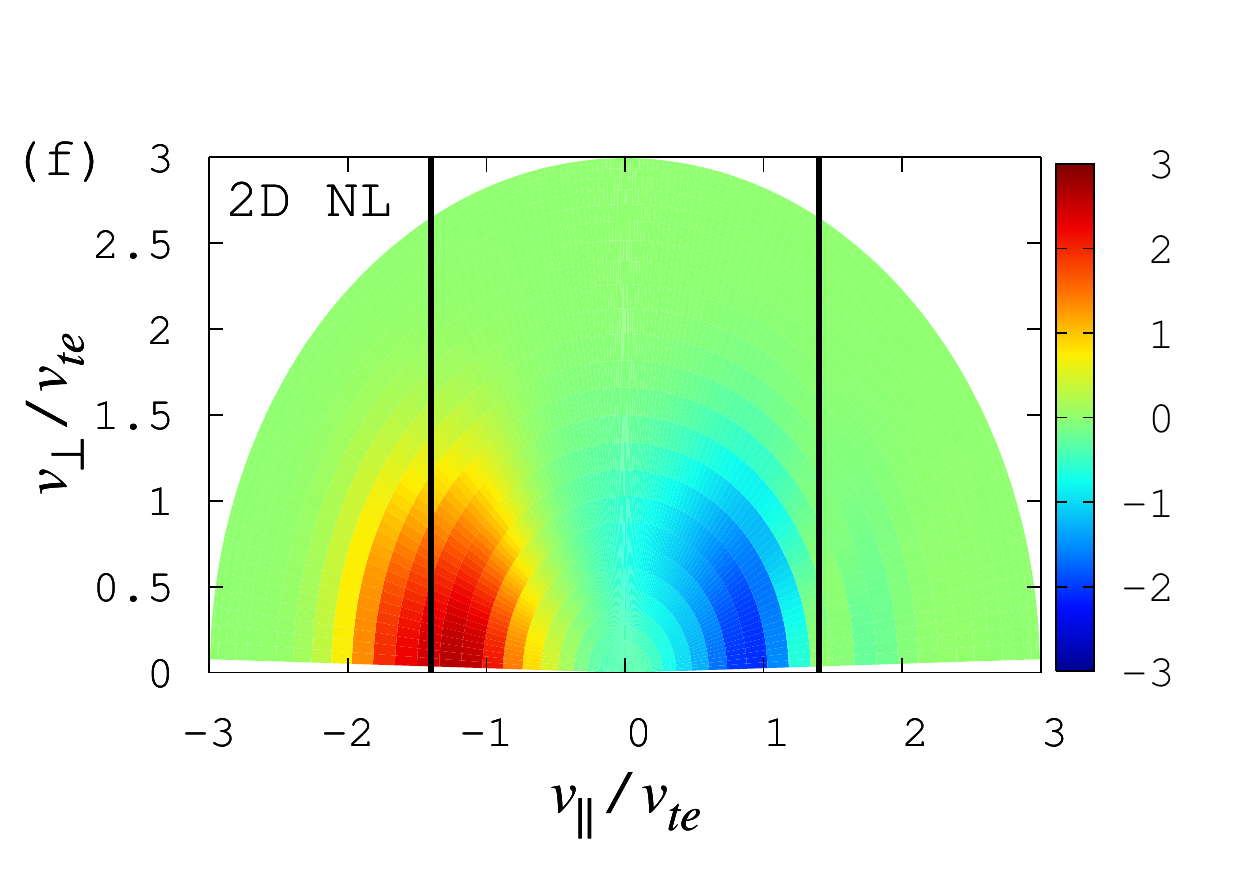}} \hfill
 \resizebox{1.8in}{!}{\includegraphics[scale=0.29]{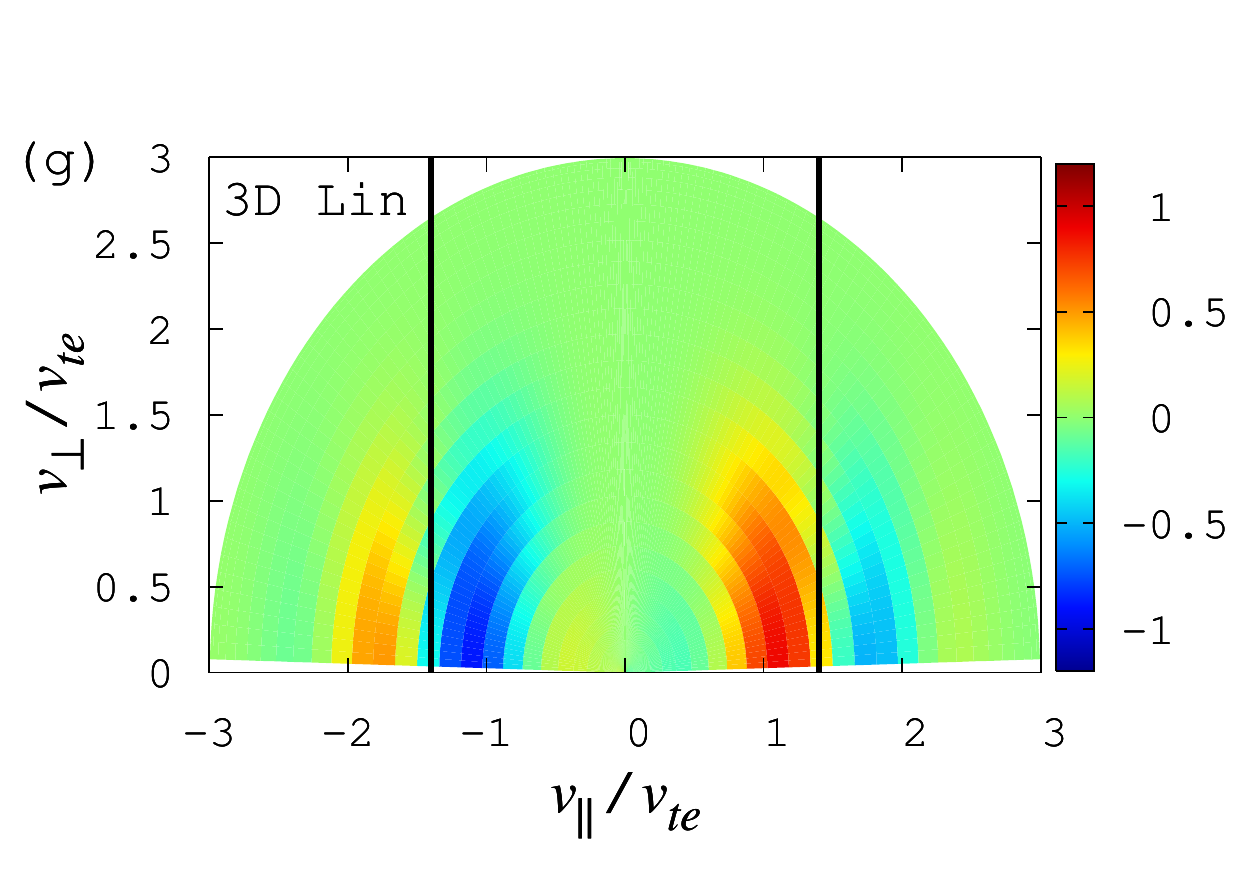}} \hfill
  \resizebox{1.8in}{!}{\includegraphics[scale=0.29]{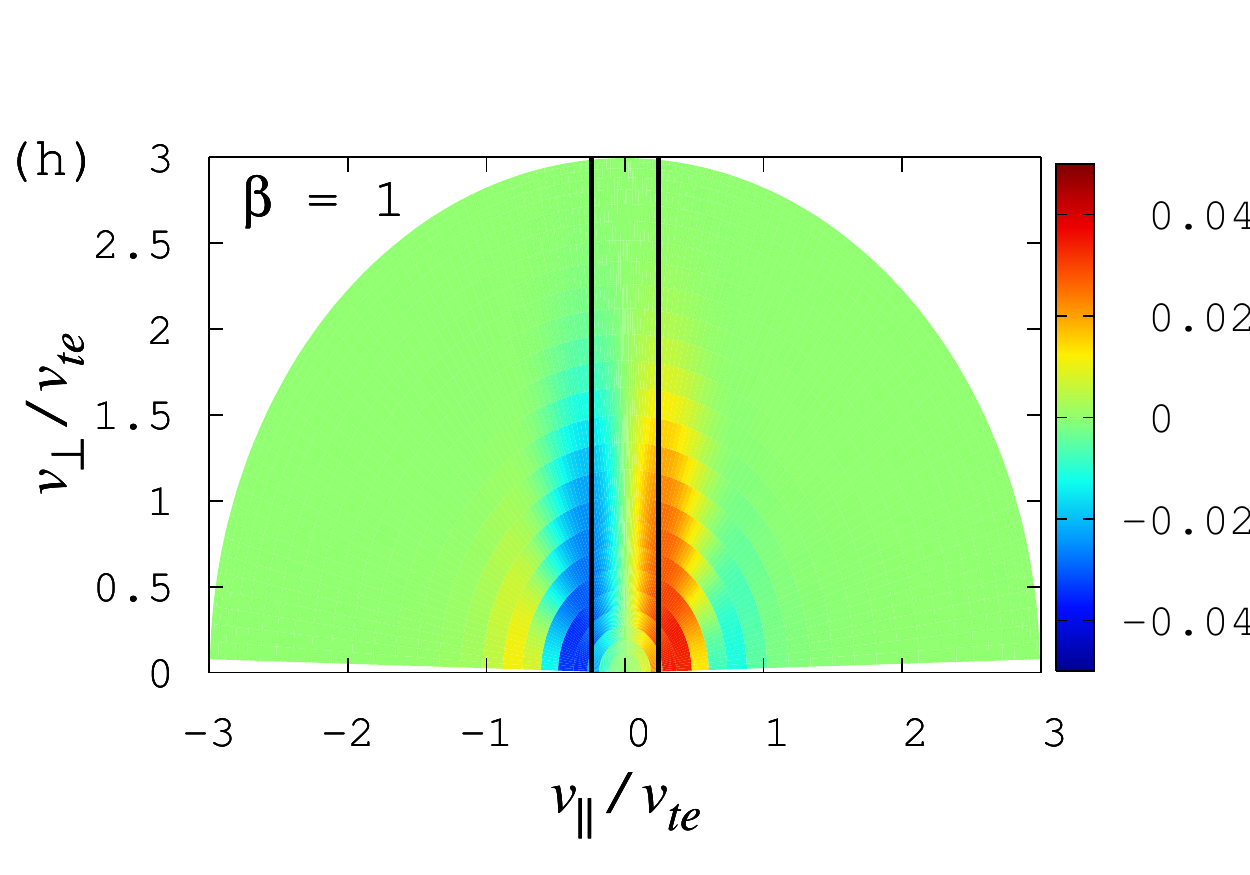}} \hfill }
\caption{ \label{fig:3d_2d_linear} Velocity-space structures in a $\beta_i=0.01$ plasma shown by plotting
    $g_e(v_\parallel,v_\perp)/\epsilon F_{0e}(v)$ for (a) OTV3D, (b)
    OTV2D and (c) 3D linear simulations for the $(1,1)$ Fourier mode
    in the $z=0$ plane, revealing only $v_\parallel$ variation. Resonant signature as an enhancement of amplitude around the resonant velocities (vertical black lines) is shown by plotting $g_e(v_\parallel,v_\perp)$ for (e) OTV3D, (f) OTV2D and (g) 3D linear cases; it is further demonstrated in (d) OTV3D and (h) 3D linear simulations for a $\beta_i=1$ plasma. Animations for OTV3D and linear simulations of both $\beta_i$ values, similar to (a) and (e), (c) and (g), are available.}
\end{figure*}
\subsection{Landau damping}
Landau damping \citep{Landau:1946} is the mechanism by which particles absorb energy from parallel electric fields of waves in collisionless plasmas. Waves and particles interact strongly near the resonant velocity $v_r$ when the Landau resonance condition, $\omega$ - $k_\parallel v_\parallel$ = 0, is satisfied, leading to an enhanced amplitude in the distribution function near $v_r$. The parallel electric field responsible for the interaction naturally generates velocity-space structure varying in $v_\parallel$. Furthermore, for \Alfven waves at $k_\perp \rho_e < 1$, linear gyrokinetic theory predicts that fluctuations in $g_e(v_\parallel,v_\perp)/F_{0e}(v)$ via Landau damping vary only in $v_\parallel$, with little variation in $v_\perp$ \citep{Howes:2006}. Both an enhanced amplitude near the resonance condition and only $v_\parallel$ variation are the two key characteristics of Landau damping to be focused on below.

\subsection{Velocity-space structures indicating Landau damping}
The velocity-space structures observed in both the 2D
and 3D cases agree with the characteristics of Landau damping that are also exhibited in linear simulations in which \Alfven waves are known to be Landau damped.

In \figref{fig:3d_2d_linear}, for the (a) OTV3D and (b) OTV2D cases,
we plot $g_e(v_\parallel,v_\perp)/\epsilon F_{0e}(v)$ (the amplitude of $g_e$ relative to $F_{0e}$, normalized by the gyrokinetic epsilson) for the
perpendicular Fourier mode in the $z=0$ plane, $(k_x \rho_i,k_y \rho_i,z)=(1,1,0)$, at the peak of the electron collisional
dissipation rate at $t/\tau_0\simeq$ 0.7. The resonant velocities for
this \Alfven mode are $|v_r|=|\omega/k_\parallel| \simeq 
1.4 v_{te}$ (vertical black lines), where $\omega/k_\parallel$ is given by linear kinetic theory. Note that this mode receives
energy strictly via the turbulent nonlinear interactions that transfer
energy from the larger scale initial modes. Plotted in (c) is the
velocity-space characteristic of Landau damping in the $z=0$ plane from a linear 3D simulation of two counterpropagating \Alfven waves with
$(k_x \rho_i,k_y \rho_i,k_zL_z)=(1,1,\pm 1)$, in which the waves are
verified to Landau damp at the rate predicted by linear kinetic
theory.  In all of these plots, since $F_{0e}(v)$ is small at large $v$, division by $F_{0e}(v)$ emphasizes structures at larger $v$, thus enabling small-amplitude fluctuations near the tail of the distribution to be more clearly seen. This is good for revealing the overall profile of the variation. In the plots (a)-(c), OTV3D, OTV2D and 3D linear simulations share the same strictly $v_\parallel$ structure, consistent with the characteristic of Landau damping. The variation in OTV2D develops shorter scales, closely resembling those in OTV3D, at a later time, consistent with the slower evolution of OTV2D (see \figref{fig:os23otv}).

Another observed characteristic in the turbulence simulations is an enhanced amplitude around $v_r$. This can be examined by plotting just $g_e(v_\parallel,v_\perp)$ (instead of the relative amplitude). In \figref{fig:3d_2d_linear}, we plot $g_e(v_\parallel,v_\perp)$ for (e) OTV3D, (f) OTV2D and (g) 3D linear simulations, showing an enhancement of fluctuations near $v_r$. The specific profile varies from case to case, but they all share the same characteristic of exhibiting an increase in amplitude around $v_r$.

To confirm the robustness of the resonance signature, identical OTV3D
and 3D linear simulations but with $\beta_i=1$, in which, the resonance will occur at a much lower $|v_r| \simeq 0.24 v_{te}$, for the same (1,1) mode, were performed. An enchanced amplitude is observed at the predicted lower $v_r$ in $g_e(v_\parallel,v_\perp)$ from (d) OTV3D
and (h) 3D linear simulations, confirming the presence of Landau resonance. Supplemental movies further illustrate the persistent signature around $v_r$ throughout the time evolution of OTV3D and 3D linear simulations.

The velocity-space structures in \figref{fig:3d_2d_linear} provide a
novel means for the characterization of the physical mechanism
responsible for the collisionless damping of fluctuations in kinetic
plasma turbulence. We find that both 2D and 3D simulations of strong
plasma turbulence develop velocity-space characteristics indicative of Landau damping, namely, only $v_\parallel$ variation and an enhanced amplitude around the resonant condition. This development of velocity-space structures facilitates the subsequent collisional
dissipation of the non-thermal energy. Furthermore, the velocity-space characteristics in the nonlinear simulations appears very similar to that of
\emph{linear} Landau damping, suggesting the kinetic dissipation is
largely linear in nature, supporting a recent theoretical prediction
\citep{Howes:2015b}.

Let us mention two additional points. First, the entropy
cascade \citep{Schekochihin:2009,Tatsuno:2009} is not expected to play
a significant role for electrons in the transfer of energy to small
velocity-space scales when $k_\perp \rho_e < 1$, and indeed we find
little of the $v_\perp$ variation predicted for this process. Second,
a current sheet develops during the merging of the initial double
vortices, and multiple thin current sheets arise later in both the 2D
and 3D simulations (\figref{fig:os23otv}). Despite the development and
dissipation of these coherent structures, the velocity-space structures appears to be consistent with Landau damping,
supporting speculation that Landau damping dominates dissipation of
current sheets in collisionless 3D kinetic \Alfven wave turbulence
\citep{TenBarge:2013a}, and consistent with the interpretation of solar
wind observations \citep{Sahraoui:2009} and evidence of Landau damping in 2D simulations of magnetic reconnection \citep{Numata:2015}.

\subsection{Landau Damping in 2D?}
For a 2D simulation with no variation
along the equilibrium magnetic field $\V{B}_0=B_0\hat{\mathbf{z}}$,
one may naively expect that Landau damping is prohibited.  However,
when there is an in-plane component such that the total field is
$\mathbf{B} = B_0 \zhat + \delta B_\perp\hat{\mathbf{x}}$, the
resonant denominator responsible for Landau damping becomes $\omega -
k_xv_\parallel\sin\theta$, where $\sin\theta= \delta
B_\perp/|\mathbf{B}|$. The frequency of linear \Alfven waves
propagating on the 2D plane is also modified to become $\omega =
k_xv_A\sin\theta$. The trigonometric correction factors out, leaving a
2D resonant condition $v_\parallel \sim v_A$ that is essentially
unchanged from the 3D case.  This explains how Landau damping can potentially play an important role in energy dissipation in both 2D and 3D kinetic
turbulence simulations. In 2D hybrid simulations of \Alfvenic turbulence, cyclotron resonance is also observed \citep{Hellinger:2015}.


\section{Conclusions}
Results from 2D and 3D nonlinear gyrokinetic simulations of plasma turbulence at $\beta_i \ll 1$ are presented. Qualitatively similar evolution of the energy flow from turbulent fluctuations to electron heat is observed in both 2D and 3D cases. Quantitatively, the nonlinear energy transfer and subsequent dissipation is substantially slower in the 2D case. In addition, the development of electron velocity-space structures is examined to characterize the nature of the dissipation process. We found that throughout the time evolution of both the 2D and 3D nonlinear simulations, fluctuations enhanced near the resonant velocity with only
$v_\parallel$ structure, characteristics shared by the
linear Landau damping of \Alfven waves, are generated. This suggests the continual occurrence of Landau damping throughout the dissipation process and also indicates that this kinetic damping mechanism is essentially linear in nature
\citep{Howes:2015b}. This is further supported by complementary 3D nonlinear and linear kinetic simulations using $\beta_i$ = 1, which are consistent with the role of Landau damping in 3D Landau-fluid simulations of \Alfvenic turbulence \citep{Passot:2014}. Despite naive expectations that Landau damping
is ineffective in 2D, the action of Landau damping in 2D simulations
is explained theoretically by a common trigonometric correction factor
appearing in both the resonant denominator and the linear wave
frequency, resulting in an essentially unchanged resonance
condition. These results provide new information on the dynamics of distribution functions during the dissipation process and are consistent with previous finding that Landau
damping likely plays a dominant role in energy transfer in plasma
turbulence, even when current sheets develop and dissipate, as
observed in 3D simulations of kinetic \Alfven wave turbulence
\citep{TenBarge:2013a,TenBarge:2013b} and 2D simulations of
strong-guide-field magnetic reconnection \citep{Numata:2015}. Future studies require investigation of the connection between the turbulent fields and plasma particles in velocity space to determine the energy transfer between the fields and particles.

\acknowledgments
Supported by NSF CAREER Award AGS-1054061, NASA grant NNX10AC91G, NSF
grant AGS-1331355, and US DOE grant DEFG0293ER54197. This work used
the Extreme Science and Engineering Discovery Environment (XSEDE),
which is supported by NSF grant ACI-1053575.

%

\begin{thebibliography}{}
\expandafter\ifx\csname natexlab\endcsname\relax\def\natexlab#1{#1}\fi

\bibitem[{{Abel} {et~al.}(2008){Abel}, {Barnes}, {Cowley}, {Dorland}, \&
  {Schekochihin}}]{Abel:2008}
{Abel}, I.~G., {Barnes}, M., {Cowley}, S.~C., {Dorland}, W., \& {Schekochihin},
  A.~A. 2008, Phys.~Plasmas, 15, 122509

\bibitem[{{Barnes} {et~al.}(2009){Barnes}, {Abel}, {Dorland}, {Ernst},
  {Hammett}, {Ricci}, {Rogers}, {Schekochihin}, \& {Tatsuno}}]{Barnes:2009}
{Barnes}, M., {Abel}, I.~G., {Dorland}, W., {et~al.} 2009, Phys.~Plasmas, 16,
  072107

\bibitem[{{Brizard} \& {Hahm}(2007)}]{Brizard:2007}
{Brizard}, A.~J., \& {Hahm}, T.~S. 2007, Rev. Mod. Phys., 79, 421

\bibitem[{{Che} {et~al.}(2014){Che}, {Goldstein}, \& {Vi{\~n}as}}]{Che:2014}
{Che}, H., {Goldstein}, M.~L., \& {Vi{\~n}as}, A.~F. 2014, Phys.~Rev.~Lett.,
  112, 061101

\bibitem[{{Dahlburg} \& {Picone}(1989)}]{Dahlburg:1989}
{Dahlburg}, R.~B., \& {Picone}, J.~M. 1989, Phys.~Fluids B, 1, 2153

\bibitem[{Franci {et~al.}(2015)Franci, Verdini, Matteini, Landi, \&
  Hellinger}]{Franci:2015}
Franci, L., Verdini, A., Matteini, L., Landi, S., \& Hellinger, P. 2015, The
  Astrophysical Journal Letters, 804, L39

\bibitem[{{Frieman} \& {Chen}(1982)}]{Frieman:1982}
{Frieman}, E.~A., \& {Chen}, L. 1982, Phys.~Fluids, 25, 502

\bibitem[{{Gary} {et~al.}(2008){Gary}, {Saito}, \& {Li}}]{Gary:2008}
{Gary}, S.~P., {Saito}, S., \& {Li}, H. 2008, Geophys.~Res.~Lett., 35, 2104

\bibitem[{Goldreich \& Sridhar(1995)}]{Goldreich:1995}
Goldreich, P., \& Sridhar, S. 1995, Astrophys.~J., 438, 763

\bibitem[{{Grauer} \& {Marliani}(2000)}]{Grauer:2000}
{Grauer}, R., \& {Marliani}, C. 2000, Phys.~Rev.~Lett., 84, 4850

\bibitem[{{Haynes} {et~al.}(2014){Haynes}, {Burgess}, \&
  {Camporeale}}]{Haynes:2014}
{Haynes}, C.~T., {Burgess}, D., \& {Camporeale}, E. 2014, Astrophys.~J., 783,
  38

\bibitem[{Hellinger {et~al.}(2015)Hellinger, Matteini, Landi, Verdini, Franci,
  \& Trávníček}]{Hellinger:2015}
Hellinger, P., Matteini, L., Landi, S., {et~al.} 2015, The Astrophysical
  Journal Letters, 811, L32

\bibitem[{Howes(2015)}]{Howes:2015b}
Howes, G.~G. 2015, Philosophical Transactions of the Royal Society of London A:
  Mathematical, Physical and Engineering Sciences, 373, 20140145

\bibitem[{{Howes}(2015)}]{Howes:2015a}
{Howes}, G.~G. 2015, Journal of Plasma Physics, 81, 3203

\bibitem[{{Howes} {et~al.}(2006){Howes}, {Cowley}, {Dorland}, {Hammett},
  {Quataert}, \& {Schekochihin}}]{Howes:2006}
{Howes}, G.~G., {Cowley}, S.~C., {Dorland}, W., {et~al.} 2006, Astrophys.~J.,
  651, 590

\bibitem[{{Howes} {et~al.}(2008){Howes}, {Dorland}, {Cowley}, {Hammett},
  {Quataert}, {Schekochihin}, \& {Tatsuno}}]{Howes:2008a}
{Howes}, G.~G., {Dorland}, W., {Cowley}, S.~C., {et~al.} 2008,
  Phys.~Rev.~Lett., 100, 065004

\bibitem[{{Howes} {et~al.}(2014){Howes}, {Klein}, \& {TenBarge}}]{Howes:2014a}
{Howes}, G.~G., {Klein}, K.~G., \& {TenBarge}, J.~M. 2014, Astrophys.~J., 789,
  106

\bibitem[{Howes {et~al.}(2011)Howes, TenBarge, Dorland, Quataert, Schekochihin,
  Numata, \& Tatsuno}]{Howes:2011a}
Howes, G.~G., TenBarge, J.~M., Dorland, W., {et~al.} 2011, Phys.~Rev.~Lett.,
  107, 035004

\bibitem[{Landau(1946)}]{Landau:1946}
Landau, L.~D. 1946, Zh. Eksp. Teor. Fiz., 16, 574

\bibitem[{{Markovskii} \& {Vasquez}(2013)}]{Markovskii:2013}
{Markovskii}, S.~A., \& {Vasquez}, B.~J. 2013, Astrophys.~J., 768, 62

\bibitem[{{Mininni} {et~al.}(2006){Mininni}, {Pouquet}, \&
  {Montgomery}}]{Mininni:2006}
{Mininni}, P.~D., {Pouquet}, A.~G., \& {Montgomery}, D.~C. 2006,
  Phys.~Rev.~Lett., 97, 244503

\bibitem[{{Narita} {et~al.}(2014){Narita}, {Comisel}, \&
  {Motschmann}}]{Narita:2014}
{Narita}, Y., {Comisel}, H., \& {Motschmann}, U. 2014, Frontiers in Phys., 2,
  13

\bibitem[{{Numata} {et~al.}(2010){Numata}, {Howes}, {Tatsuno}, {Barnes}, \&
  {Dorland}}]{Numata:2010}
{Numata}, R., {Howes}, G.~G., {Tatsuno}, T., {Barnes}, M., \& {Dorland}, W.
  2010, J.~Comp.~Phys., 229, 9347

\bibitem[{{Numata} \& {Loureiro}(2015)}]{Numata:2015}
{Numata}, R., \& {Loureiro}, N.~F. 2015, Journal of Plasma Physics, 81, 3001

\bibitem[{{Orszag} \& {Tang}(1979)}]{Orszag:1979}
{Orszag}, S.~A., \& {Tang}, C.-M. 1979, J.~Fluid Mech., 90, 129

\bibitem[{{Parashar} {et~al.}(2009){Parashar}, {Shay}, {Cassak}, \&
  {Matthaeus}}]{Parashar:2009}
{Parashar}, T.~N., {Shay}, M.~A., {Cassak}, P.~A., \& {Matthaeus}, W.~H. 2009,
  Phys.~Plasmas, 16, 032310

\bibitem[{{Parashar} {et~al.}(2014){Parashar}, {Vasquez}, \&
  {Markovskii}}]{Parashar:2014b}
{Parashar}, T.~N., {Vasquez}, B.~J., \& {Markovskii}, S.~A. 2014, Physics of
  Plasmas, 21, 022301

\bibitem[{Passot {et~al.}(2014)Passot, Henri, Laveder, \& Sulem}]{Passot:2014}
Passot, T., Henri, P., Laveder, D., \& Sulem, P.-L. 2014, The European Physical
  Journal D, 68, 1

\bibitem[{{Perrone} {et~al.}(2013){Perrone}, {Valentini}, {Servidio}, {Dalena},
  \& {Veltri}}]{Perrone:2013}
{Perrone}, D., {Valentini}, F., {Servidio}, S., {Dalena}, S., \& {Veltri}, P.
  2013, Astrophys.~J., 762, 99

\bibitem[{{Picone} \& {Dahlburg}(1991)}]{Picone:1991}
{Picone}, J.~M., \& {Dahlburg}, R.~B. 1991, Phys.~Fluids B, 3, 29

\bibitem[{{Politano} {et~al.}(1989){Politano}, {Pouquet}, \&
  {Sulem}}]{Politano:1989}
{Politano}, H., {Pouquet}, A., \& {Sulem}, P.~L. 1989, Phys.~Fluids B, 1, 2330

\bibitem[{{Politano} {et~al.}(1995){Politano}, {Pouquet}, \&
  {Sulem}}]{Politano:1995b}
---. 1995, Physics of Plasmas, 2, 2931

\bibitem[{{Sahraoui} {et~al.}(2009){Sahraoui}, {Goldstein}, {Robert}, \&
  {Khotyaintsev}}]{Sahraoui:2009}
{Sahraoui}, F., {Goldstein}, M.~L., {Robert}, P., \& {Khotyaintsev}, Y.~V.
  2009, Phys.~Rev.~Lett., 102, 231102

\bibitem[{{Schekochihin} {et~al.}(2009){Schekochihin}, {Cowley}, {Dorland},
  {Hammett}, {Howes}, {Quataert}, \& {Tatsuno}}]{Schekochihin:2009}
{Schekochihin}, A.~A., {Cowley}, S.~C., {Dorland}, W., {et~al.} 2009,
  Astrophys.~J.~Supp., 182, 310

\bibitem[{{Servidio} {et~al.}(2012){Servidio}, {Valentini}, {Califano}, \&
  {Veltri}}]{Servidio:2012}
{Servidio}, S., {Valentini}, F., {Califano}, F., \& {Veltri}, P. 2012,
  Phys.~Rev.~Lett., 108, 045001

\bibitem[{Servidio {et~al.}(2015)Servidio, Valentini, Perrone, Greco, Califano,
  Matthaeus, \& Veltri}]{Servidio:2015}
Servidio, S., Valentini, F., Perrone, D., {et~al.} 2015, Journal of Plasma
  Physics, 81, 1

\bibitem[{{Tatsuno} {et~al.}(2009){Tatsuno}, {Schekochihin}, {Dorland},
  {Plunk}, {Barnes}, {Cowley}, \& {Howes}}]{Tatsuno:2009}
{Tatsuno}, T., {Schekochihin}, A.~A., {Dorland}, W., {et~al.} 2009,
  Phys.~Rev.~Lett., 103, 015003

\bibitem[{TenBarge {et~al.}(2014)TenBarge, Daughton, Karimabadi, Howes, \&
  Dorland}]{TenBarge:2014b}
TenBarge, J.~M., Daughton, W., Karimabadi, H., Howes, G.~G., \& Dorland, W.
  2014, Phys.~Plasmas, 21, 020708

\bibitem[{{TenBarge} \& {Howes}(2013)}]{TenBarge:2013a}
{TenBarge}, J.~M., \& {Howes}, G.~G. 2013, Astrophys.~J.~Lett., 771, L27

\bibitem[{{TenBarge} {et~al.}(2013){TenBarge}, {Howes}, \&
  {Dorland}}]{TenBarge:2013b}
{TenBarge}, J.~M., {Howes}, G.~G., \& {Dorland}, W. 2013, Astrophys.~J., 774,
  139

\bibitem[{{Thompson} \& {Lysak}(1996)}]{Thompson:1996}
{Thompson}, B.~J., \& {Lysak}, R.~L. 1996, \jgr, 101, 5359

\bibitem[{{Verscharen} {et~al.}(2012){Verscharen}, {Marsch}, {Motschmann}, \&
  {M{\"u}ller}}]{Verscharen:2012a}
{Verscharen}, D., {Marsch}, E., {Motschmann}, U., \& {M{\"u}ller}, J. 2012,
  Phys.~Plasmas, 19, 022305

\bibitem[{Wan {et~al.}(2015)Wan, Matthaeus, Roytershteyn, Karimabadi, Parashar,
  Wu, \& Shay}]{Wan:2015}
Wan, M., Matthaeus, W.~H., Roytershteyn, V., {et~al.} 2015, Phys. Rev. Lett.,
  114, 175002

\bibitem[{{Wan} {et~al.}(2012){Wan}, {Matthaeus}, {Karimabadi}, {Roytershteyn},
  {Shay}, {Wu}, {Daughton}, {Loring}, \& {Chapman}}]{Wan:2012}
{Wan}, M., {Matthaeus}, W.~H., {Karimabadi}, H., {et~al.} 2012,
  Phys.~Rev.~Lett., 109, 195001

\bibitem[{{Wu} {et~al.}(2013{\natexlab{a}}){Wu}, {Wan}, {Matthaeus}, {Shay}, \&
  {Swisdak}}]{Wu:2013b}
{Wu}, P., {Wan}, M., {Matthaeus}, W.~H., {Shay}, M.~A., \& {Swisdak}, M.
  2013{\natexlab{a}}, Phys.~Rev.~Lett., 111, 121105

\bibitem[{{Wu} {et~al.}(2013{\natexlab{b}}){Wu}, {Perri}, {Osman}, {Wan},
  {Matthaeus}, {Shay}, {Goldstein}, {Karimabadi}, \& {Chapman}}]{Wu:2013a}
{Wu}, P., {Perri}, S., {Osman}, K., {et~al.} 2013{\natexlab{b}},
  Astrophys.~J.~Lett., 763, L30

\end{thebibliography}

\end{document}